\documentclass[article,onecolumn,showpacs]{revtex4}
\usepackage{amsmath}
\usepackage{amssymb}
\usepackage{color}
\usepackage{verbatim}
\usepackage{array}
\usepackage{graphicx}
\usepackage{graphics,psfrag}
\usepackage{graphicx,psfrag}
\usepackage{epsfig}
\usepackage{multirow}

\newcommand{\ds}{\displaystyle}
\newcommand{\ddsum}[1]{{\displaystyle \sum_{ #1 }}}

\newcommand{\supercomas}[1]{``#1''}

\def\bra#1{\mathinner{\langle{#1}|}}

\def\ket#1{\mathinner{|{#1}\rangle}}

\begin{document}
\preprint{\hfill\parbox[b]{0.3\hsize}{ }}

\title{Photon-photon polarization correlations as a tool for studying 
parity non-conservation in heliumlike Uranium}

\author{Filippo Fratini$^{1,2}$\footnote{
E-mail address: fratini@physi.uni-heidelberg.de}, Sergiy
Trotsenko$^{2,3}$, Stanislav Tashenov$^{1}$, Thomas St\"ohlker$^{1,2,3}$ and Andrey Surzhykov$^{1,2}$ } \affiliation {\it
$^1$
Physikalisches Institut, Heidelberg University, D-69120 Heidelberg, Germany \\
$^2$
GSI Helmholtzzentrum f\"ur Schwerionenforschung (GSI), D-64291 Darmstadt, Germany\\
$^3$
Helmholtz-Institut Jena, D-07743 Jena, Germany }

\begin{abstract}
Due to electron--nucleus weak interaction, atomic bound states with different parities turn out to be mixed.
We discuss a prospect for measuring the mixing parameter between
the nearly degenerate metastable states $1s_{1/2}2s_{1/2} \, : \, J=0$ and $1s_{1/2}2p_{1/2} \, : \, J=0$ in heliumlike Uranium.
Our analysis is based on the polarization properties of the photons emitted in the two--photon decays of such states.
\end{abstract}


\maketitle

\section{Introduction} 
\label{sec:intr}
Parity non-conservation (PNC) had been at first theoretically proposed by Lee and Yang in 1956 
in order to find a way out of the so--called \supercomas{$\tau - \gamma$ puzzle} \cite{PR104, PR100}. The next year,
Wu and collaborators observed an asymmetry in nuclear beta decay ascribed to
parity non-conservation in weak processes \cite{PR105}, and, subsequently, Lee and Yang have been forthwith awarded with the Nobel Prize.
Many later experiments in nuclear and high energy physics confirmed
parity violation in weak interactions and
precisely recorded weak charge and other related parameters \cite{PL77B, PRC34, PR297, PR107}.
Although with some initial controversies, 
the \supercomas{$\tau - \gamma$ puzzle} was also solved out by understanding that 
both $\tau$ and $\gamma$ were two decay channels of the same parent particle, 
known today as the charged kaon $K^+$ \cite{PR106, AIP300}.
In contrast to nuclear and high energy physics, fewer experiments have been carried out in atomic physics to measure 
weak interaction's properties. In fact,
the conflicting results of the early Bismuth experiments in the '70 \cite{PRL39, PRL39L, PL85B, PRL46} spread the conviction that nothing
fundamentally useful could have ever been extracted from atomic physics experiments. Nonetheless, renewed interest 
on the subject rose again in the late '80 and '90 
and led to the successful measurements of the weak charge $Q_w$ 
and related parameters in atomic Cesium \cite{PRL61, S272, PRL102, PRL82, PRL85, PRA62}, Thallium \cite{PRL74}, Lead \cite{PRL71}
and Ytterium \cite{PRL103}.
On the theoretical side, starting from the early work of Curtis-Michel \cite{PR138B}, several investigations 
of PNC have been made in the context of neutral atoms \cite{PRA37}, few--electron ions \cite{PL48B, PRA63}
and muonic atoms \cite{PLB256, PR118}. In all the proposed studies, the little role played by PNC effects together with 
the need of precise measurements have been highlighted.

\medskip

Parity violation in atomic physics is mainly caused by the exchange of the $Z^0$ boson between atomic electrons and quarks in the nucleus.
All atomic states become mixtures consisting mainly of
the state they are usually assigned, together with a small percentage of states possessing the opposite parity.
It has been discussed some time ago the prospect for measuring the mixing between the states
$1s_{1/2}2s_{1/2} : J = 0$ and $1s_{1/2}2p_{1/2} : J = 0$ 
in heliumlike Uranium by inducing 
a resonant parity violating $E1E1$ transition between them \cite{PRA40}. The authors of the paper 
concluded that the proposed measurement
was not feasible with the then available technology, while, nowadays, is under consideration at GSI facility in
Darmstadt (Germany). With the same goal, some years later Dunford proposed 
an analysis based on the circular polarization asymmetry of one of the photons emitted in the two-photon decay of $1s_{1/2}2p_{1/2} : J = 0$ 
state \cite{PRA54}.
As the author concluded, the calculations therein performed were not enough to assess whether or not the polarization asymmetry could
lead to useful parity experiments.
With the same intent and similar method, we propose 
another route based on photon polarization properties, for
the experimental determination of the mixing parameter between the states $1s_{1/2}2s_{1/2} : J = 0$ and $1s_{1/2}2p_{1/2} : J = 0$ in U$^{90+}$
(in the following, these two states will be briefly called $2^1S_0$ and $2^3P_0$ respectively).
The different polarization properties of the photons emitted in the two--photon decays of such states 
suggest a way to discriminate the decays and, thereby, to measure the mixing parameter between the states. 
However, 
the prospect presents some technical difficulties, 
widely discussed in the text, that make the experimental realization a challenge with the current state of art technologies.

\medskip

The paper is structured as follows.
Sec.~\ref{sec:HeUranium} describes the salient characteristics of the first excited 
states of U$^{90+}$, Sec.~\ref{sec:Geom} shows the geometry we refer to, while Sec.~\ref{sec:Theory} 
describes the two--photon transition amplitude and the employed atomic model. 
The polarization--polarization correlation, that
is the function which denotes the probability of detection 
of photons with certain polarizations, will be also discussed in details. In Sec.~\ref{sec:RaD},
results are shown
for such correlation function, emphasizing the role played by parity mixing terms. 
With results in hand, the experimental set--up for the prospect is then explained and widely discussed. Finally, a brief summary is
given at the end.
\begin{figure}[t]
\center
\includegraphics[width=.4\textwidth]{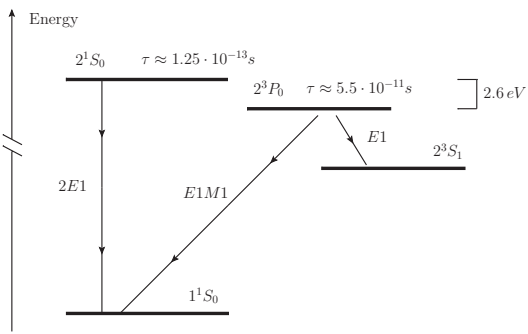}
\caption{Level scheme of few low energetic states in U$^{90+}$.}
\label{fig:scheme}
\end{figure}

\section{Heliumlike Uranium ion}
\label{sec:HeUranium}
Heliumlike Uranium ion represents a very suitable candidate 
for studying PNC, due to the fact that the states 
$2^3P_0$ and $2^1S_0$  are separated by an energy difference of only few electronvolts \cite{PRA53,PRA81}, out of a
total binding energy of order $100$ KeV.
Fig.~\ref{fig:scheme} shows the scheme of the first levels of U$^{90+}$ \cite{PRL57}.
$2^3P_0$ state has negative parity and lifetime of about $\sim10^{-11}$ sec, 
while $2^1S_0$ state has positive parity and
shorter lifetime of about $\sim10^{-13}$ sec. 
Although $2^3P_0$ can decay by single photon emission into 
$2^3S_1$, both $2^3P_0$ and $2^1S_0$ decay exclusively by two--photon decay to the ground state, owing to angular momentum conservation.
Due to weak interaction between electrons and nucleus, 
$2^3P_0$ acquires a small admixture of $2^1S_0$ and vice versa. Since
the size of the parity mixing depends inversely on the energy difference between the mixed states \cite{PRA40},
both $2^3P_0$ and $2^1S_0$ do not get any other considerable PNC contribution from any other state.
More explicitly, at the first order in perturbation theory, 
the \supercomas{{\it true}} $\ket{2^3P_0}$ and $\ket{2^1S_0}$ states can be written as \cite{PRA54}
\begin{equation}
\begin{array}{l c l}
\ds \ket{\tilde{2^3P_0}}&\approx&\ket{2^3P_0}+\eta\ket{2^1S_0}  \\
\ds \ket{\tilde{2^1S_0}}&\approx&\ket{2^1S_0}+\eta\ket{2^3P_0}   ~,
\end{array}
\label{eq:statesmixed}
\end{equation}
where 
the tilde notation is here and henceforth used to denote \supercomas{true} states, 
in order to differentiate them from the bare theoretical Dirac states which will be denoted without the tilde.
The mixing parameter $\eta$ in Eq.~\ref{eq:statesmixed} is given by
\begin{equation}
\begin{array}{l c l}
\ds\eta&=&\ds\frac{\bra{2^3P_0}
\hat{H}_W
\ket{2^1S_0}}{\Delta E_{PS}}~,
\end{array}
\label{eq:eta}
\end{equation}
where $\Delta E_{PS}$ is the energy difference between $2^3P_0$ and $2^1S_0$, while $\hat{H}_W$ is the operator
for the nuclear-spin-independent weak interaction \cite{PRA54}.
Up to a very good approximation, 
we will neglect any parity mixing effect 
in any state with the exception of $2^3P_0$ and $2^1S_0$. 
Among the low energetic states in U$^{90+}$, only these two
have in fact energies near enough to determine 
a sizeable mixing parameter between them.\\

\section{Geometry}
\label{sec:Geom}
\begin{figure}[t!]
\centering
\includegraphics[width=10cm, height=6cm]{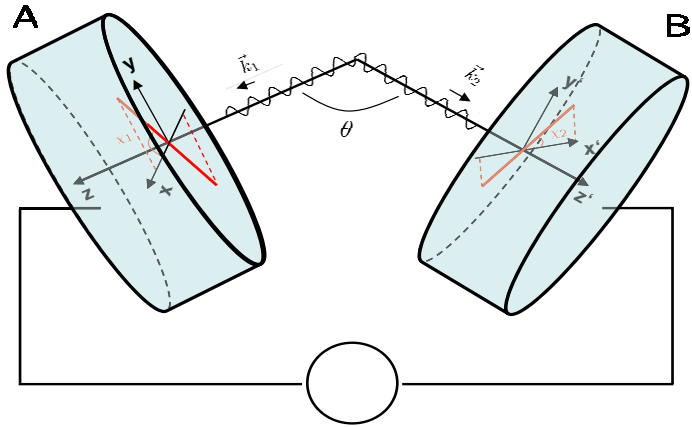}
\caption{Geometry for the two--photon decay. 
The $z\;(z')$ axis is adopted along the propagation direction of the first (second)
photon, the $x$ axis is chosen such that $xz$ is the reaction plane, while
the $y$ axis is coincident with $y'$. The definitions of the opening angle and of the angles which define the 
detectors' transmission
axes are also displayed.}
\label{fig:angleDEF}
\end{figure}

The geometry we want to adopt for the prospect is displayed in Fig.~\ref{fig:angleDEF}.
We define one local system of reference for each emitted photon. The axes definitions are as follows:
the propagation direction of the first (second) photon is adopted as $z$ ($z'$) while
the angle between the photons' directions (opening angle) is called
$\theta$. The $x$-axis is fixed such that the plane defined by the two photons' directions (reaction plane) is the $xz$-plane.\\
Using standard notation, we call \supercomas{photon polarization plane}
the plane which is orthogonal to the photon's direction with the origin of coordinate axes located at the position where the photon is detected.
As displayed in Fig.~\ref{fig:angleDEF}, the $A(B)$ detector measures the
linear polarization of the first (second) photon along the transmission axis
defined by the angle $\chi_{1(2)}$ in the polarization plane. 
The detectors are thought to work as polarizer filters: 
whenever a photon hits any one of them, such detector gives off or not a \supercomas{click}, which would respectively
indicate that the photon has been measured as having its polarization along the direction $\chi_{1,2}$ or $\chi_{1,2}+90^{\circ}$.\\
Finally, we define the sharing parameter $f$ as the fraction of energy carried away by the first
photon:
\begin{equation}
   f=\frac{\omega_1}{E_i-E_f }=1-\frac{\omega_2}{E_i-E_f } \, ,
\end{equation}
where $E_{i,f}$ are the energies of the initial and final ionic states while
$\omega_{1,2}$ are the recorded energies of the first and second photon. Energy conservation
has been used in the last step of the above equation.

\section{Theory}
\label{sec:Theory}
On the purpose of measuring the parameter $\eta$ in Eq.~(\ref{eq:eta}), we propose to prepare $U^{90+}$ in $\tilde{2^3P_0}$ state.
The efficiency of such preparation is here assumed to be $100\%$. 
The so prepared $\tilde{2^3P_0}$ state will decay either into $2^3S_1$ or 
into the ground state, as extensively
displayed in Fig.~\ref{fig:scheme}. The two--photon decay channel $\tilde{2^3P_0}\to1^1S_0$, in which we are interested,
can be easily selected out in experiments by requiring a (two-detector) coincidence measurement. 
The amplitude for such process can be obtained in second order perturbation theory and reads \cite{PRA24}
\begin{equation}
\begin{array}{l}
\ds\mathcal{M}^{\lambda_1\lambda_2}(\tilde{2^3P_0}\to 1^1S_0)= 
\ddsum{\nu}\!\!\!\!\!\!\!\!\int
 \left[
\frac{ \bra{1^1S_0} \vec\alpha\cdot\vec u_{\lambda_1}^*e^{-i\vec k_1\cdot \vec r}
\ket{\nu}\bra{\nu} \vec\alpha\cdot\vec u_{\lambda_2}^*e^{-i\vec k_2\cdot \vec r\,'}\ket{\tilde{2^3P_0}}}{E_{\nu}
-E_i+\omega_2 } 
+ \right. \\
\qquad \qquad \qquad \qquad \ds \qquad \left. \frac{ \bra{1^1S_0} \vec\alpha\cdot\vec u_{\lambda_2}^*e^{-i\vec k_2\cdot \vec r}
\ket{\nu}\bra{\nu} \vec\alpha\cdot\vec u_{\lambda_1}^*e^{-i\vec
k_1\cdot \vec r\,'}\ket{\tilde{2^3P_0}}}{E_{\nu} -E_i+\omega_1 } \right] ~.
\end{array}
\label{eq:Mfi}
\end{equation}
Here $\vec \alpha$ is the vector of Dirac matrices while 
the symbol $\ds\ddsum{\nu}\!\!\!\!\!\!\!\!\int $ stands for
both a summation over the discrete and an integration over the continuum part of the ionic spectrum. In addition, 
$E_{\nu}$ is the energy
of the intermediate electronic state $\ket{\nu}$, while $\vec k_{1,2}$ and $\vec{u}_{\lambda_{1,2}}$ denote
respectively the linear momentum and the polarization vector of the first, second photon. The latter directly depends on the photons helicities
$\lambda_{1,2}=-1,1$.
By introducing (\ref{eq:statesmixed}) into (\ref{eq:Mfi}), the amplitude splits up into two terms,
\begin{equation}
\mathcal{M}^{\lambda_1\lambda_2}(\tilde{2^3P_0}\to1^1S_0)\approx \mathcal{M}^{\lambda_1\lambda_2}(2^3P_0\to1^1S_0) + 
\eta \mathcal{M}^{\lambda_1\lambda_2}(2^1S_0\to1^1S_0) ~.
\label{eq:firstM}
\end{equation}
In order to suggest any experiment whose goal is the measurement of the mixing parameter $\eta$, we should be first able to theoretically
discriminate the two amplitudes of the right-hand side of Eq.~(\ref{eq:firstM}).
The key point of the prospect is that such discrimination can be obtained 
by studying the photons' polarization properties contained in those amplitudes. 
It has been recently showed that, in case that 
nearly equal energy is shared between the photons,
the two--photon decay $2^3P_0\to 1^1S_0$ is characterized by photon linear polarizations which are exclusively {\it orthogonal}
to each other (linear polarizations of the first, second
photon are detected, correspondingly, 
along the axes $x,y'$ or $y,x'$), while the two--photon decay $2^1S_0\to 1^1S_0$ is 
characterized by photon linear polarizations 
which are 
exclusively {\it parallel}
to each other (linear polarizations of the first, second
photon are detected, correspondingly, along the axes $x,x'$ or $y,y'$) \cite{Hyperfine, QuantumInf}.
While the first assertion is true independently of the 
opening angle $\theta$, the second one holds only in case the photons 
are recorded either collinearly  or back--to--back ($\theta=0^{\circ},180^{\circ}$). However, as it will be evident in the following,
the linear polarizations of photons emitted in $2^1S_0\to 1^1S_0$ decay can be considered parallel
in the whole intervals $0^{\circ}\le \theta\lesssim 2^{\circ}$ and $178^{\circ}\lesssim \theta\le180^{\circ}$, 
due to the fact that the (orthogonal) corrections to the polarization state
are negligible in that region,
even for the delicate problem under consideration.
As a matter of fact,
for the case $0^{\circ}\le \theta\lesssim 2^{\circ}$ (or $178^{\circ}\lesssim \theta\le180^{\circ}$) and $f=0.5$,  it can be demonstrated that 
the polarization state of the two photons emitted in consequence of the decay of
the prepared $\tilde{2^3P_0}$ state can be simply described by the
ket vector \cite{QuantumInf} 
\begin{equation}
\begin{array}{l}
\ds \ket{\Psi} = O_{f,Z,\theta}^{PS} \, \big(\ket{xy} + \ket{yx}\big) + 
\eta \, O_{f,Z,\theta}^{SS} \, \big(\ket{xx} + \ket{yy}\big) \, ,
\end{array}
\label{eq:state}
\end{equation}
where $\Big|O_{f,Z,\theta}^{PS}\Big|^2$ is the probability of
detecting the emitted photons with polarizations along $\chi_1=0^{\circ}$, $\chi_2=90^{\circ}$
or $\chi_1=90^{\circ}$, $\chi_2=0^{\circ}$ while
$\Big|O_{f,Z,\theta}^{SS}\Big|^2$ is the probability of
detecting the photons with polarizations along $\chi_1=0^{\circ}$, $\chi_2=0^{\circ}$
or $\chi_1=90^{\circ}$, $\chi_2=90^{\circ}$.
Both $O_{f,Z,\theta}^{PS}$ and $O_{f,Z,\theta}^{SS}$
contain the dependence on the energy sharing 
parameter $f$, the atomic number $Z$ and the opening angle $\theta$
given respectively by the amplitudes $\mathcal{M}^{\lambda_1,\lambda_2}(2^3P_0\to 1^1S_0)$ and 
$\mathcal{M}^{\lambda_1,\lambda_2}(2^1S_0\to 1^1S_0)$.

\medskip

In order to inspect the polarization properties of the photons emitted in the two-photon decay 
of the $\tilde{2^3P_0}$ state, 
we define the polarization--polarization correlation function, which is
the physical quantity we 
mean to investigate. 
Such function is given by \cite{Hyperfine}
\begin{equation}
\begin{array}{c c c}
   \ds \Phi_{\chi_1,\chi_2}^f(\theta) &=& \ds
   \frac{\mathcal{N}^2}{4 (2J_i +1)} \,
   \sum\limits\limits_{\substack{\lambda_1 \lambda'_1 \\ \lambda_2 \lambda'_2}}
   {\rm e}^{i (\lambda_1 - \lambda'_1) \chi_1} \,
   {\rm e}^{i (\lambda_2 - \lambda'_2) \chi_2} \,
   \mathcal{M}^{\lambda_1\lambda_2}(i\to f)
   \mathcal{M}^{\lambda_1'\lambda_2' *}(i\to f) \, ,
\end{array}   
\label{eq:corrfunction}
\end{equation}
where $J_i$ is the total angular momentum of the initial ionic state.
Thus $\Phi$ represents the normalized probability density of measuring, in coincidence, two photons
with well--defined wave vectors $\vec k_1$, $\vec k_2$
and with certain linear polarizations which are characterized by the angles $\chi_1$, $\chi_2$ with respect 
to the reaction plane (see Fig.~\ref{fig:angleDEF} for details concerning the used notation).
The normalization constant $\mathcal{N}$ is chosen such that $1/\mathcal{N}^2$ is the sum of the probability densities of the four 
independent polarization outcomes 
$\chi_{1,2}$ = 0$^{\circ}$, 90$^{\circ}$.\\
In order to complete the theoretical background needed for the
prospect, we conclude this section by explaining the model we use for the calculations.
The description of two--electron ions is indeed a theoretical challenge of the current state of research in atomic physics. 
The method of relativistic finite basis sets, for instance, has been shown 
to be valid and efficient in order to obtain highly accurate 
calculations of the two-photon E1M1 decay rate from the $2^3P_0$ state \cite{NIMPRB9}.
Alternatively, the salient 
characteristics of heavy heliumlike ions can be described 
by the Independent Particle Model (IPM), which is the model used here for the calculations. 
Although such model treats the electrons as independent particles bound to the nucleus (the nuclear Coulomb attraction
is assumed to be much stronger than the electron--electron repulsion),
it takes the Pauli principle into account. 
Moreover, this model allows a drastic simplification of the two--electron amplitude which appears in Eq.~(\ref{eq:firstM}),
allowing to reduce it 
to a summation over one--electron amplitudes \cite{PRA81_2}. Although the calculation of the latter quantity is itself 
a challenging theoretical problem, 
several methods have been successfully proposed in the past decades to precisely 
perform it \cite{PRA23, PRA77}.
The tool we adopt here for its calculation is
the relativistic Dirac-Coulomb Green function.
For details regarding this approach, useful information can be found in Refs.~\cite{PRA71, JPA24}.\\
The results showed in Sec.~\ref{sec:RaD} are obtained by taking into account the full multipoles contribution of the photons' fields.
Finally, the {\it effective} nuclear charge used for the computation is $Z=91.275$.
Such shrewdness accounts for the electromagnetic screening that one electron
makes on the other one, allowing for a basic electron--electron interaction. 
\begin{figure}
\psfrag{= tildePtoS}[bl][b][0.6]{$\!\!\!\!\!\!\!\!\!\!\!\!\!\!\!= \tilde{2^3P_0}\to 1^1S_0$}
\psfrag{= PtoS}[bl][b][0.6]{$\!\!\!\!\!\!\!\!\!\!= 2^3P_0\to 1^1S_0$}
\psfrag{= StoS}[bl][b][0.6]{$\!\!\!\!\!\!\!\!\!\!= 2^1S_0\to 1^1S_0$}
\center
\includegraphics[width=8cm, height=6cm]{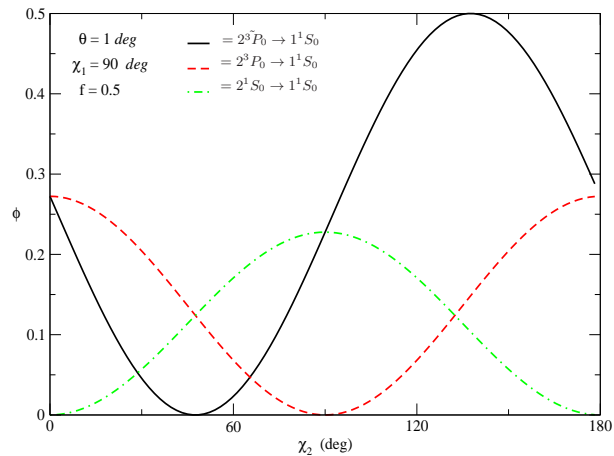}
\caption{
(Color online) Polarization--polarization correlation function (\ref{eq:corrfunction}) for the $\tilde{2^3P_0}\to1^1S_0$ two--photon
decay of heliumlike Uranium ion. 
The contribution of the different amplitudes in Eq.(\ref{eq:firstM}) are separately displayed.
The dashed (red) line and the dot--dashed (green) line represent respectively
the $P\to S$ and the $S\to S$ contribution to the correlation function, while the solid (black) line denotes the total $\tilde{P}\to S$
correlation function. Here $\theta$ is the opening angle, $f$ is the 
photons' sharing energy and $\chi_{1,2}$ are the linear polarization 
angles at which the first, second detectors' transmission axes are set (cfr. Fig.~\ref{fig:angleDEF} for details).
}
\label{fig:PNCTheor}
\end{figure}
\section{Prospect: results and discussion}
\label{sec:RaD}
After having explained the theory at the base of our prospect as well as the model we use, 
we are now 
ready to present concretely the proposal.\\
In order to measure the mixing parameter $\eta$ in Eq.~(\ref{eq:eta}),
we propose, as previously mentioned, to prepare the Uranium ion U$^{90+}$ in $\tilde{2^3P_{0}}$ state, to
place two polarization detectors at a fixed position in the reaction plane and to use them as 
polarizer filters. While one of the two detectors will be kept
at a fixed orientation (fixed transmission axis), the transmission axis of the other one will be continuously 
rotated to record the correlation function (\ref{eq:corrfunction}) for different photons polarization configurations.
In order to suggest a workable experimental scenario, 
we must inevitably look for opening angle and energy 
values which enable $\eta\, O_{f,Z,\theta}^{SS}$ to be comparable with 
$O_{f,Z,\theta}^{PS}$, in Eq.~(\ref{eq:state}).
In other words, since $\eta$ is considerably small, we must
find a configuration where the amplitude $\mathcal{M}^{\lambda_1,\lambda_2}(2^3P_0 \to 1^1S_0)$ is small in comparison with 
$\mathcal{M}^{\lambda_1,\lambda_2}(2^1S_0 \to 1^1S_0)$.
On this purpose, it has been showed that
the decay rate for the $2^3P_0\to 1^1S_0$ transition is strongly suppressed for 
photons' opening 
angle $0\le\theta\lesssim 2^{\circ}$ and equal energy sharing, whereas, for the same configuration, the decay rate $2^1S_0\to 1^1S_0$
gets (almost) its maximum value \cite{PRA81_2, PRA69}. 
Choosing small values of $\theta$ and equal energy sharing will also ensure that the 
different amplitudes in Eq.~(\ref{eq:firstM}) will determine different photons polarization outcomes (as
remarked in Sec.~\ref{sec:Theory}), which is decisive for the scope of the prospect.
An optimal configuration for our purposes can be found, for instance,
at $\theta=1^{\circ}$ and f=0.5.
For such a configuration, the coefficients $O_{f,Z,\theta}^{PS}$ and $O_{f,Z,\theta}^{SS}$,
which compose the ket vector (\ref{eq:state}), assume respectively the values  -8.49$\cdot$10$^{-11}$ and
4.43$\cdot$10$^{-5}$.
The correlation function $\Phi$ related to such polarization state can be easily calculated: 
\begin{equation}
\begin{array}{l c l}
\ds \Phi_{\chi_1, \chi_2}^{f=0.5}(\theta=1^{\circ}) &=&\mathcal{N}^{\,2}\,\Big[ -8.49\cdot10^{-11} \, 
\big(  \cos\chi_1\sin\chi_2 + \sin\chi_1\cos\chi_2 \big) + 
\eta \, 4.43\cdot10^{-5} \, \big( \cos\chi_1\cos\chi_2 + \sin\chi_1\sin\chi_2    \big) \Big]^2\, .
\end{array}
\label{eq:PPcorr2}
\end{equation}
We draw the above function in Fig.~\ref{fig:PNCTheor},
where $\chi_1$ has been arbitrarily set to $90^{\circ}$, for a better visualization, 
while $\eta$ has been fixed to the predicted theoretical value $1.75\cdot10^{-6}$, which can be obtained
by correcting the value obtained in Ref.~\cite{PRA40} with the precise calculation of the $2^1S_0 - 2^3P_0$ energy gap showed 
in Ref.~\cite{PRA81}.
The different contributions of the two addends in Eq.~(\ref{eq:firstM}) are separately displayed, as well as
the total correlation function.
We can easily notice that the \supercomas{parity allowed} ($\ket{xy} + \ket{yx}$) and 
\supercomas{parity forbidden} ($\ket{xx} + \ket{yy}$) 
components of the photon polarization state have approximately the same magnitude.
In concordance with Ref.~\cite{Hyperfine}, it 
can be seen in the figure, as well as from Eq.~(\ref{eq:PPcorr2}), that
the amplitudes $\mathcal{M}^{\lambda_1,\lambda_2}(2^1S_0\to1^1S_0)$ and $\mathcal{M}^{\lambda_1,\lambda_2}(2^3P_0\to1^1S_0)$
determine respectively the probability of detecting parallel and orthogonal linearly polarized photons.
In an ideal experiment, we could therefore 
scan the function $\Phi$ over the whole or part of the domain $\chi_{1,2}\in[0,180^{\circ}]$, in order to be then able
to determine the parameter $\eta$ by fitting the measured polarization correlation with the ($\eta$ dependent) function (\ref{eq:PPcorr2}).

\medskip

The proposal is based 
on the fact that, for $f\to 0.5$ and $\theta\to0$, the transition $2^3P_0\to1^1S_0$ (model-independently) vanishes. 
If we considered
the two-photon transition $\tilde{2^1S_0}\to1^1S_0$, it can be easily seen from 
Eqs.~(\ref{eq:statesmixed}) and (\ref{eq:Mfi}), that
the amplitude for that process would turn out to be equal to $(\ref{eq:firstM})$,
with the replacement 
$\mathcal{M}^{\lambda_1,\lambda_2}(2^1S_0\to1^1S_0)\leftrightarrow\mathcal{M}^{\lambda_1,\lambda_2}(2^3P_0\to1^1S_0)$.
Since, unfortunately, there exist no geometry for which the transition $2^1S_0\to1^1S_0$ is suppressed, 
the polarization of the emitted photons would be completely dominated by the \supercomas{parity allowed} component that,
in that case, would be ($\ket{xx} + \ket{yy}$).
An initial preparation 
of the $\tilde{2^1S_0}$ state, therefore,
although easier from an experimental point of view \cite{PRA74,PRL104},
would {\it not} give rise to the interference pattern shown in Fig.~\ref{fig:PNCTheor}, for any given geometry.\\
Moreover,
the amplitude $\mathcal{M}^{\lambda_1,\lambda_2}(2^1S_0\to1^1S_0)$ is approximately one order of magnitude 
larger than $\mathcal{M}^{\lambda_1,\lambda_2}(2^3P_0\to1^1S_0)$, as can be seen from the states' lifetimes displayed in Fig.~\ref{fig:scheme}. 
This fact represents an advantage for studying PNC effects in $\tilde{2^3P_0}\to1^1S_0$ rather than in 
$\tilde{2^1S_0}\to1^1S_0$, since such difference
compensates partially for the small value of the mixing parameter $\eta$ in Eq.~(\ref{eq:firstM})
and so helps the two addends in the same equation to be comparable.\\

\medskip

Although the suggested settings $f=0.5$ and $\theta=1^{\circ}$ ensure, as needed, that the $2^3P_0\to1^1S_0$ channel is strongly suppressed,
they determine at the same time a challenging arrangement for the experimental investigation of the prospect. 
Specifically, because of the required small opening angle $\theta$,
the two X-ray photon detectors would have to be placed
at a relatively long distance from the source of radiation and thus the detection efficiency 
would be substantially decreased.
An additional hindrance lies in the fact that the 
polarizations of both photons have to be measured at equal energy sharing ($f=0.5$).
For the case of Uranium, this fact would imply that each photon has about 50~keV (rest frame) energy.
The polarization resolved experiments in this X-ray energy regime are nowadays normally
performed by using Compton polarimeters \cite{TNS52a, TNS52b, RSI79, JI5, RSI67, PRL97}. By 
selecting events recorded in coincidence which have the desired (Compton) scattering angle,
such polarimeters can be used to measure the polarization state of the photon pair.
The selection of events can however increase considerably the statistical uncertainty.\\
Further experimental difficulties for the realization of the prospect might arise from the 
angle-energy resolution needed to record the interference pattern shown in Fig~\ref{fig:PNCTheor}.
The $P\to S$ channel rises fast, glossing over the other $S\to S$ channel in which we are interested, 
as soon as we depart from the (exact) theoretical proposed configuration $f=0.5$, $\theta= 1^{\circ}$.
In other words, slightly different angle-energy settings would bring about a completely different polarization--polarization
correlation function with respect to (\ref{eq:PPcorr2}). As a matter of fact, the opening angle and energy
resolutions needed, in order to select events for which the correlation function does not change approximately shape, would 
be, according to our calculations, respectively 0.5 degrees and 5 electronvolts.
Even though the required angle resolution may be nowadays achieved, the energy resolution
needed is approximately three orders of magnitude higher than the available resolution in current Compton polarimeters. 
A possible way to overcome the energy resolution limitation would be to use the so-called 
absorption edge technique~\cite{Stoehlker97}. In this technique the photons pass through an apsorption foil. 
The K-shell absorption edge of the foil atoms serves as a photon energy filter. 
The photons with the energy below the K-shell photoionization energy will have a significantly higher 
transmission probability than the photons with the higher energies. Since in the proposed experimental scheme 
both of the entangled photons have the same energy, one foil can be used as the energy filter for both of the photons. 
By adjusting the ion beam velocity the photon energy can be Doppler-tuned such that it is less than 5~eV below the K-edge. 
A Compton scattering polarimeter behind the absorption foil can be then used for the polarization analysis of the transmitted photons. 
Another possible experimental approach would involve high energy resolution calorimeters and a Rayleigh scattering polarimetry 
technique~\cite{Tashenov09}. Here the energy of the Rayleigh scattered photon and its scattering direction could be measured 
with high resolution by an array of x-ray calorimeters. Such arrays are currently being developed~\cite{Porter04,Silver05} 
and likely to reach the required energy resolution at the energy of 50~keV in the near future. 

\medskip

The small (expected) value of the mixing coefficient $\eta$ is certainly at the base
of the technical difficulties explained above.
A way to lighten such difficulties might be represented, for instance, by 
selecting a suitable isotope of U$^{90+}$. In virtue of the fact that the energy gap between 
$2^1S_0$ and $2^3P_0$ states varies slightly by changing the mass number of the ion \cite{PRA81}, the mixing 
of the two states itself would depend on the considered isotope (cfr. Eq.~(\ref{eq:eta})).
In particular, by choosing an isotope of Uranium whose mass number is smaller than 238, we would be able to 
increase the mixing parameter of the two states up to a factor $\approx$1.6. However, besides the technical difficulties
related to the radioactive properties that the chosen isotope might show,
such improvement would be anyway not enough to bring considerable advantages to the prospect.
\section{Summary}
In summary, a prospect for measuring the parity mixing parameter between the states $1s_{1/2}2s_{1/2} : J=0$ and
$1s_{1/2}2p_{1/2} : J=0$ in heliumlike Uranium has been presented. The core of the prospect lies on
the discernment of the two--photon decay of such states by using the polarization properties of the emitted photons. 
Within relativistic second--order perturbation theory and
the independent particle model, we explored 
the polarization--polarization correlation function of the photon pair for a chosen angle-energy configuration in which 
the role played by parity mixing terms is highlighted.
Within the suggested settings, 
the presence of parity mixing contributions changes quantitatively {\it and} qualitatively
the shape of the correlation function in the overall domain $0\leq\chi_{1,2}\leq180^{\circ}$.
Such changes could be in principle measured in a polarization-angle resolved experiment.
However, the prospect presents some technical difficulties, discussed in the text, 
which currently hamper its experimental investigation.
The theoretical analysis which has been carried out on the polarization properties of the emitted photons,  
may be also used as a side study for any other experimental investigations of PNC effects in atoms or ions.

\begin{acknowledgments}
A.S. and F.F. acknowledge support by the Helmholtz Gemeinschaft under the project VH-NG-421 and from 
the Deutscher Akademischer Austauschdienst (DAAD) under the Project No. 0813006. 
S.T. acknowledges support by the German Research Foundation (DFG) within the Emmy Noether 
program under the contract No. TA 740/1-1.  
F.F. gratefully acknowledges Renate Martin for fruitful discussions.
\end{acknowledgments}



\begin{thebibliography}{13}

\bibitem{PR104} T.~D.~Lee and C.~N.~Yang, Phys. Rev. {\bf 104}, 254 (1956).
\bibitem{PR100} G.~Harris, J.~Orear and S.~Taylor, Phys. Rev. {\bf 100}, 932 (1955).
\bibitem{PR105} C.~S.~Wu, {\it et al.}, Phys. Rev. {\bf 105}, 1413 (1957).
\bibitem{PL77B} J.~E.~Clendenin, V.~W.~Huges, {\it et al.}, Phys. Lett. {\bf 77B}, 347 (1978).
\bibitem{PRC34} J.~Lang, Th.~Maier, R.~M\"uller, {\it et al.}, Phys. Rev. C {\bf 34}, 1545 (1986).
\bibitem{PR297} B.~Desplanques, Phys. Rep. {\bf 297}, 1 (1986).
\bibitem{PR107} N.~Tanner, Phys. Rev. {\bf 107}, 1203 (1957).
\bibitem{PR106} S.~B.~Treiman and H.~W.~Wyld, Phys. Rev. {\bf 106}, 1320 (1957).
\bibitem{AIP300} R.~H.~Dalitz, AIP Conf. Proc. {\bf 300}, 141 (1994).
\bibitem{PRL39} P.~E.~G.~Baird, M.~W.~S.~M.~Brimicombe, R.~G.~Hunt, G.~J.~Roberts, P.~G.~H.~Sandars and D.~N.~Stacey, 
Phys. Rev. Lett. {\bf 39}, 798 (1977).
\bibitem{PRL39L} L.~L.~Lewis, J.~H.~Hollister, D.~C.~Soreide, E.~G.~Lindahl and E.~N.~Fortson, Phys. Rev. Lett. {\bf 39}, 795 (1977).
\bibitem{PL85B} L.~M.~Barkov and M.~S.~Zolotorev, Phys. Lett. {\bf 85B}, 308 (1979).
\bibitem{PRL46} J.~H.~Hollister, G.~R.~Apperson, L.~L.~Lewis, T.~P.~Emmons, T.~G.~Vold and E.~N.~Fortson, Phys. Rev. Lett. {\bf 46}, 643 (1981).
\bibitem{PRL61} M.~C.~Noecker, B.~P.~Masterson and C.~E.~Wieman, Phys. Rev. Lett. {\bf 61}, 310 (1988).
\bibitem{S272} C.~S.~Wood, {\it et al.}, Science {\bf 275}, 1759 (1997).
\bibitem{PRL102} S.~G.~Porsev, K.~Beloy and A.~Derevianko, Phys. Rev. Lett. {\bf 102}, 181601 (2009).
\bibitem{PRL82} S.~C.~Bennett and C.~E.~Wieman, Phys. Rev. Lett. {\bf 82}, 2484 (1999).
\bibitem{PRL85} A.~Derevianko. Phys. Rev. Lett. {\bf 85}, 1618 (2000).
\bibitem{PRA62} V.~A.~Dzuba and V.~V.~Flambaum, Phys. Rev. A {\bf 62}, 052101 (2000).
\bibitem{PRL74} P.~A.~Vetter, D.~M.~Meekhof, P.~K.~Majumder, S.~K.~Lamoreaux and E.~N.~Fortson, Phys. Rev. Lett. {\bf 74}, 2658 (1995).
\bibitem{PRL71} D.~M.~Meekhof, P.~Vetter, P.~K.~Majumder, S.~K.~Lamoreaux and E.~N.~Fortson, Phys. Rev. Lett. {\bf 71}, 3442 (1993).
\bibitem{PRL103} K.~Tsigutkin, D.~Dounas-Frazer, A.~Family, J.~E.~Stalnaker, V.~V.~Yashchuk and D.~Budker, Phys. Rev. Lett. {\bf 103}, 071601 (2009).
\bibitem{PR138B} F.~Courtis-Michel, Phys. Rev. {\bf 138}, B408 (1965).
\bibitem{PRA37} W.~R.~Johnson, S.~A.~Blundell, Z.~W.~Liu and J.~Sapirstein, Phys. Rev. A {\bf 37}, 1395 (1988).
\bibitem{PL48B} M.~A.~Bouchiat and C.~C.~Bouchiat, Phys. Lett. {\bf 48B}, 111 (1974).
\bibitem{PRA63} L.~N.~Labzowsky, A.~V.~Nefiodov, G.~Plunien, G.~Soff, R.~Marrus and D.~Liesen, Phys. Rev. A {\bf 63}, 054105 (2001).
\bibitem{PLB256} P.~Langacker, Phys. Lett. {\bf 256B}, 277 (1990).
\bibitem{PR118} J.~Missimer and L.~M.~Simons, Phys. Rep. {\bf 118}, 179 (1985).
\bibitem{PRA40} A.~Sch\"afer, G.~Soff, P.~Indelicato, B.~M\"uller and W.~Greiner, Phys. Rev. A {\bf 40}, 7362 (1989).
\bibitem{PRA54} R.~W.~Dunford, Phys. Rev. A {\bf 54}, 3820 (1996).
%
\bibitem{PRA53} M.~Maul, A.~Sch\"afer, W.~Greiner and P.~Indelicato, Phys. Rev. A {\bf 53}, 3915 (1996).
\bibitem{PRA81} F.~Ferro, A.~Artemyev, Th.~St\"ohlker and A.~Surzhykov, Phys. Rev. A {\bf 81}, 062503 (2010).
\bibitem{PRL57} C.~T.~Munger and H.~Gould, Phys. Rev. Lett. {\bf 57}, 2927 (1986).
\bibitem{PRA24} S.P.Goldman and G.W.F.Drake, Phys.Rev.A {\bf 24}, 183 (1981).
\bibitem{Hyperfine} F.~Fratini and A.~Surzhykov, arXiv:1006.4799, accepted for publication in Hyp. Int. .
\bibitem{QuantumInf} F.~Fratini, M.~C.~Tichy, T.~Jahrsetz, A.~Buchleitner, S.~Fritzsche and A.~Surzhykov, 
                     Phys. Rev. A {\bf 83}, 032506 (2011).
\bibitem{NIMPRB9} G.~W.~F.~Drake, Nucl. Instrum. and Methods B {\bf 9}, 465-470 (1985).
\bibitem{PRA81_2} A.~Surzhykov, A.~Volotka, F.~Fratini, J.~P.~Santos, P.~Indelicato, G.~Plunien, 
                  Th.~St\"ohlker and S.~Fritzsche, Phys. Rev. A {\bf 81}, 042510 (2010).
\bibitem{PRA23} G.~W.~F.~Drake and S.~P.~Goldman, Phys. Rev. A {\bf 23}, 2093 (1981).
\bibitem{PRA77} T.~Radtke, A.~Surzhykov, S.~Fritzsche, Phys. Rev. A {\bf 77}, 0022507 (2008).
\bibitem{JPA24} R.~A.~Swainson and G.~W.~Drake, J. Phys. A {\bf 24}, 95 (1991).
\bibitem{PRA71} A.~Surzhykov, P.~Koval and S.~Fritzsche, Phys. Rev. A {\bf 71}, 0022509 (2005).
\bibitem{PRA69} R.~W.~Dunford, Phys. Rev. A {\bf 69}, 062502 (2004).
\bibitem{PRA74} J.~Rzadkiewicz, Th.~St\"ohlker, D.~Bana\'s, H.~F.~Beyer, F.~Bosch, C.~Brandau, 
                C.~Z.~Dong, S.~Fritzsche, A.~Gojska, A.~Gumberidze, S.~Hagmann, D.~C.~Ionescu, 
                C.~Kozhuharov, T.~Nandi, R.~Reuschl, D.~Sierpowski, U.~Spillmann, A.~Surzhykov, 
                S.~Tashenov, M.~Trassinelli and S.~Trotsenko, Phys. Rev. A {\bf 74}, 012511 (2006).
\bibitem{PRL104} S.~Trotsenko, A. Kumar, A. V. Volotka, D. Bana\'s, H. F. Beyer, H. Br\"auning, S. Fritzsche, 
                 A. Gumberidze, S. Hagmann, S. Hess, P. Jagodzi\'nski, C. Kozhuharov, R. Reuschl, S. Salem, A. Simon, 
                 U. Spillmann, M. Trassinelli, L. C. Tribedi, G. Weber, D. Winters, and Th. St\"ohlker, Phys. Rev. Lett. {\bf 104}, 033001 (2010).
\bibitem{TNS52a} D. Protic, E. L. Hull, T. Krings, and K. Vetter, IEEE Transactions on Nuclear Science {\bf 52}, 3181 (2005).
\bibitem{TNS52b} D. Protic, Th.~St\"ohlker, T. Krings, I. Mohos, and U. Spillmann, IEEE Transactions on Nuclear Science {\bf 52}, 3194 (2005).
\bibitem{RSI79}  U. Spillmann, H. Br\"auning, S. Hess, H. Beyer, Th.~St\"ohlker, J.-C. Dousse, D. Protic and T. Krings, Review of Scientific
                 Instruments {\bf 79}, 083101 (pages 8) (2008).
\bibitem{JI5} G. Weber, H. Br\"auning, S. Hess, R. M\"artin, U. Spillmann and 
                 Th.~St\"ohlker, Journal of Instrumentation {\bf 5}, C07010 (2010).
                 \bibitem{RSI67} P.~S.~Shaw, U.~Arp, A.~Henins and S.~Southworth, Rev. Sci. Instrum. {\bf 67}, 3362 (1996).
\bibitem{PRL97} S.~Tashenov, Th.~St\"ohlker, {\it et al.}, Phys. Rev. Lett. {\bf 97}, 223202 (2006).

%

\bibitem{Stoehlker97} Th.~St\"ohlker, A.~Kramer, S.~R.~Elliott, R.~E.~Marrs, J.~H.~Scofield, Phys. Rev. A {\bf 56}, 2819 (1997). 
\bibitem{Tashenov09} S.~Tashenov et al., Nucl. Inst. Meth. A {\bf 600}, 599 (2009).
\bibitem{Porter04} F.S.~Porter et al., Rev. Sci. Inst., {\bf 75} 3772 (2004).
\bibitem{Silver05} E. Silver, et al., Nucl. Instr. and Meth. A {\bf 545} 683 (2005).


\end{thebibliography}
\end{document}